\documentclass[runningheads,a4paper]{llncs}

\usepackage{times}
\usepackage{amssymb}
\usepackage{graphicx}
\usepackage{url}
\usepackage{multirow}
\usepackage{algorithm2e}
\usepackage{amssymb}
\newcommand{\keywords}[1]{\par\addvspace\baselineskip
\noindent\keywordname\enspace\ignorespaces#1}

\urldef{\mailsa}\path|frederico.oliveira@ufmt.br|

\begin{document}

% TSD 2021: Put your title here (please, use capitalization, see e.g.
% http://en.wikibooks.org/wiki/Basic_Book_Design/Capitalizing_Words_in_Titles)
\title{Evaluation of Speech Representations for MOS prediction}
%\title{CML-TTS: A non-English Multilingual Dataset for Speech Synthesis}
%\title{A Multilingual Dataset for Speech Synthesis in low-research languages}

% TSD 2021: a short form should be given in case the title is too long for the running head
\titlerunning{Evaluation of Speech Representations for MOS prediction}

% TSD 2021: Author's names. Chinese authors should write their first names(s)
% in front of their surnames. This ensures that the names appear correctly in
% If the names contain accented characters, please use escape codes
% (refer to http://en.wikibooks.org/wiki/LaTeX/Special_Characters#Escaped_codes)
\author{Frederico S. Oliveira \and Edresson Casanova \and Arnaldo Cândido Júnior \and Lucas R. S. Gris \and Anderson S. Soares \and Arlindo R. Galvão Filho}

% TSD 2021: For authors from different institutions, please use the following
% form including institution reference
% \author{Firstname1 Surname1\inst{1} \and Firstname2 Surname2 \inst{2}}

% TSD 2021: Author's names for headings. For 1-2 authors, use the following form
% \authorrunning{Frederico Oliveira and Edresson Casanova and Arnaldo C. Júnior \and Anderson S. Soares}
% TSD 2021: For more than 2 authors, please, use the following
\authorrunning{Oliveira et al.}

% TSD 2021: The authors' affiliations
\institute{UFG, Goiás - GO - Brazil \\
% TSD 2021: optional url
%\url{https://ceia.ufg.br/} \\
% TSD 2021: For authors from different institutions, add 2nd institution, etc.
% \and
% Affiliation2, Institute2, Address \\
% \url{www.website.org} \\
% \mailsb\\
}

% TSD 2021: Put all authors' names to the proceeding index (surname, first name)
\index{Oliveira, Frederico}
\index{Casanova, Edresson}
\index{Cândido Júnior, Arnaldo}
\index{Gris, Lucas}
\index{Soares, Anderson}
\index{Galvão Filho, Arlindo}

\toctitle{} \tocauthor{}

\maketitle

%
%
%	TSD 2021 SUBMISSION TEXT
%
%
\begin{abstract}
% TSD 2021:

In this paper, we evaluate feature extraction models for predicting speech quality. We also propose a model architecture to compare embeddings of supervised learning and self-supervised learning models with embeddings of speaker verification models to predict the metric MOS. Our experiments were performed on the VCC2018 dataset and a Brazilian-Portuguese dataset called BRSpeechMOS, which was created for this work. The results show that the Whisper model is appropriate in all scenarios: with both the VCC2018 and BRSpeechMOS datasets. Among the supervised and self-supervised learning models using BRSpeechMOS, Whisper-Small achieved the best linear correlation of 0.6980, and the speaker verification model, SpeakerNet, had linear correlation of 0.6963. Using VCC2018, the best supervised and self-supervised learning model, Whisper-Large, achieved linear correlation of 0.7274, and the best model speaker verification, TitaNet, achieved a linear correlation of 0.6933. Although the results of the speaker verification models are slightly lower, the SpeakerNet model has only 5M parameters, making it suitable for real-time applications, and the TitaNet model produces an embedding of size 192, the smallest among all the evaluated models. 
The experiment results are reproducible with publicly available source-code\footnote{\url{https://github.com/freds0/BSpeech-MOS-Prediction}}.

% TSD 2021: keywords, comma-separated
\end{abstract}

\section{Introduction}

The development of speech synthesis and voice conversion models has increased the need for automatic methods to evaluate the quality of generated speech. The most reliable methods among the available options rely on manual evaluation, where human evaluators are chosen to assess signal quality using a predefined numerical scale. In recent work, self-supervised learning (SSL) models have been used to predict the quality of synthesized speech. Representations obtained from models such as Wav2Vec 2.0 \cite{Baevski2020Wav2vec}, HuBERT \cite{Hsu2021Hubert}, WavLM \cite{Chen2021WavLMLS}, and TERA \cite{Liu2021Tera} have been used. These models produce high quality representations and their training requires a large amount of data.

Whisper \cite{Radford2022Whisper}, in the other hand is a for general-purpose speech recognition model based on supervised learning (SL), and it was developed with the goal of creating a robust system that generalizes well across domains, tasks and languages without relying on fine-tuning to achieve high accuracy. Whisper embeddings can be used to speech recognition, speech translation, language identification and other tasks. The authors demonstrated that training on a large and diverse supervised dataset alone can significantly enhance the robustness of speech systems. However, to date, the embeddings generated by the Whisper model have not been evaluated for their effectiveness in the task of speech quality prediction.

Speaker embeddings generated by speaker verification models (SV) offer an alternative to high-quality embeddings. Unlike the latter, speaker embeddings have a fixed size that remains constant regardless of the length of the utterance. Earlier studies, such as \cite{Wang2017does}, have examined the properties that are captured by speaker embeddings, such as the spoken content, speaker's gender, speaking rate and audio channel information. These studies have demonstrated satisfactory performance on various tasks, which has motivated further exploration of these features for predicting the quality of synthesized speech. Also, so far, the representations of SV models have not been evaluated in the speech quality prediction task.

In this paper, we propose to evaluate high-quality representations from both SL and SSL models, as well as SV representations, for the purpose of predicting the quality of synthesized speech in text-to-speech (TTS) systems. In addition, we investigate the use of these models to evaluate speech samples in a low resource dataset, in Brazilian Portuguese. Models based on SV can be an alternative to generate high quality embeddings with low computational cost, allowing the evaluation of speech quality in real time.

This paper is organized as follows: Section \ref{related_works} 
presents some prior research on automatically predicting the quality of synthesized speech. Section \ref{model_proposal} outlines the proposed model architecture developed in this study and 
Section \ref{experiments} details the experiments proposed.
Then, Section \ref{results} discusses the obtained results,
and finally, Section \ref{conclusions} presents the conclusions of this work.

\section{Related Works}\label{related_works}

Several studies have addressed the development of automatic methods for evaluating the quality of synthesized speech and have obtained results that correlate with human evaluation methods. The first pioneering work that used Deep Learning to predict quality was proposed in 2016 with the AutoMOS model by Patton et al. \cite{Patton2016Automos}. Fu et al. \cite{Fu2018Qualitynet} proposed the Quality-Net model to predict the PESQ \cite{Rix2001Pesq}, a metric which compares a degraded speech signal with a reference speech signal to provide an objective measure of the perceived voice quality by the human listener. Lo et al. \cite{Lo2019Mosnet} developed MOSNet, a improved version of Quality-Net for the MOS prediction task.

Cooper et al. \cite{Cooper2022generalization} investigate the ability of SSL models to predict speech quality in out-of-domain scenarios. With the aim of achieving this goal, the researchers conducted experiments on embedding extraction models, such as Wav2Vec 2.0 \cite{Baevski2020Wav2vec} and HuBERT \cite{Hsu2021Hubert}, and compare them to the MOSNet model. The models were trained on datasets such as the Blizzard Challenge \cite{King2016TheBlizzardChallenge} and the Voice Conversion Challenge \cite{Das2020TheVoiceConversionChallenge}, and then evaluated on the ASVSpoof 2019 \cite{Todisco2019ASVspoof} and Blizzard Challenge 2019 \cite{Wu2019TheBlizzardChallenge} datasets. 
The findings reveal that the Wav2Vec model outperforms the other evaluated models. However, evaluating without fine-tuning in a zero-shot setting proved to be challenging and resulted in a notable decrease in performance.

MOSA-Net \cite{Zezario2022Mosanet} is a cross-domain model that uses inputs from multiple domains, including spectrograms, waveforms and SSL features. According to the authors, using features from multiple domains contributes to more accurate results, and training for predicting multiple metrics outperforms the task of predicting a single one. Although the model can be adjusted to predict subjective metrics, no comparative experiments have been conducted with other models.

Tseng et al. \cite{Tseng2021} compared models for predicting MOS using embeddings generated by Wav2Vec 2.0 \cite{Baevski2020Wav2vec}, TERA \cite{Tera2020}, CAC \cite{Oord2018}, and APC \cite{Chung2019}. The authors proposed an architecture where the human's identification is the input and defines the human bias. The experiments show that the Wav2Vec model achieves the best results at the sentence and system levels. Similarly, Tseng, Kao and Lee \cite{Tseng2022DDOS} proposed DDOS, a model for MOS prediction that uses Wav2Vec 2.0 for feature extraction in conjunction with a representation of the evaluator, in order to specify the human bias. The model consists of two submodules, the regression head and the distribution head, which uses attentive pooling and DNNs to predict the score and distribution of the data. The results of the submodules are then combined to predict the MOS.

Yang et al. \cite{Yang2022} developed an framework for improving speech quality prediction by combining various SSL models, such as Wav2Vec 2.0, WavLM, HuBERT, and Data2Vec \cite{Baevski2022Data2Vec}. The framework consists of two parts: the first involves training the SSL models individually, while the second involves fusing the results of each model. The goal of the framework is to fine-tune the SSL models and enhance the accuracy of MOS prediction, treating model fusion as a technique similar to ensemble. Ragano et al. \cite{Ragano2022ACO} presented experiments comparing combining Wav2Vec 2.0 model representations with features extracted from convolutional layers, exploring different architectural combinations. Ultimately, the authors found that incorporating features extracted from convolutional layers did not improve the results.

\section{Model Proposal}\label{model_proposal}

The proposed model for evaluating the quality of synthesized speech consists of two modules: the Feature Extractor, which is responsible for extracting speech features, and the MOS Predictor, which predicts speech quality based on the extracted features. The architecture of the MOS Predictor consists of two dense blocks, ReLU activation function, and dropout. Several models are evaluated as the Feature Extractor, including SV, SL, and SSL models. The architecture of the proposed model can be seen in Figure \ref{mos_predictor_fc_paper}. Details of the selected models are given below.

\begin{figure}[!ht]
\centering
\includegraphics[width=4cm]{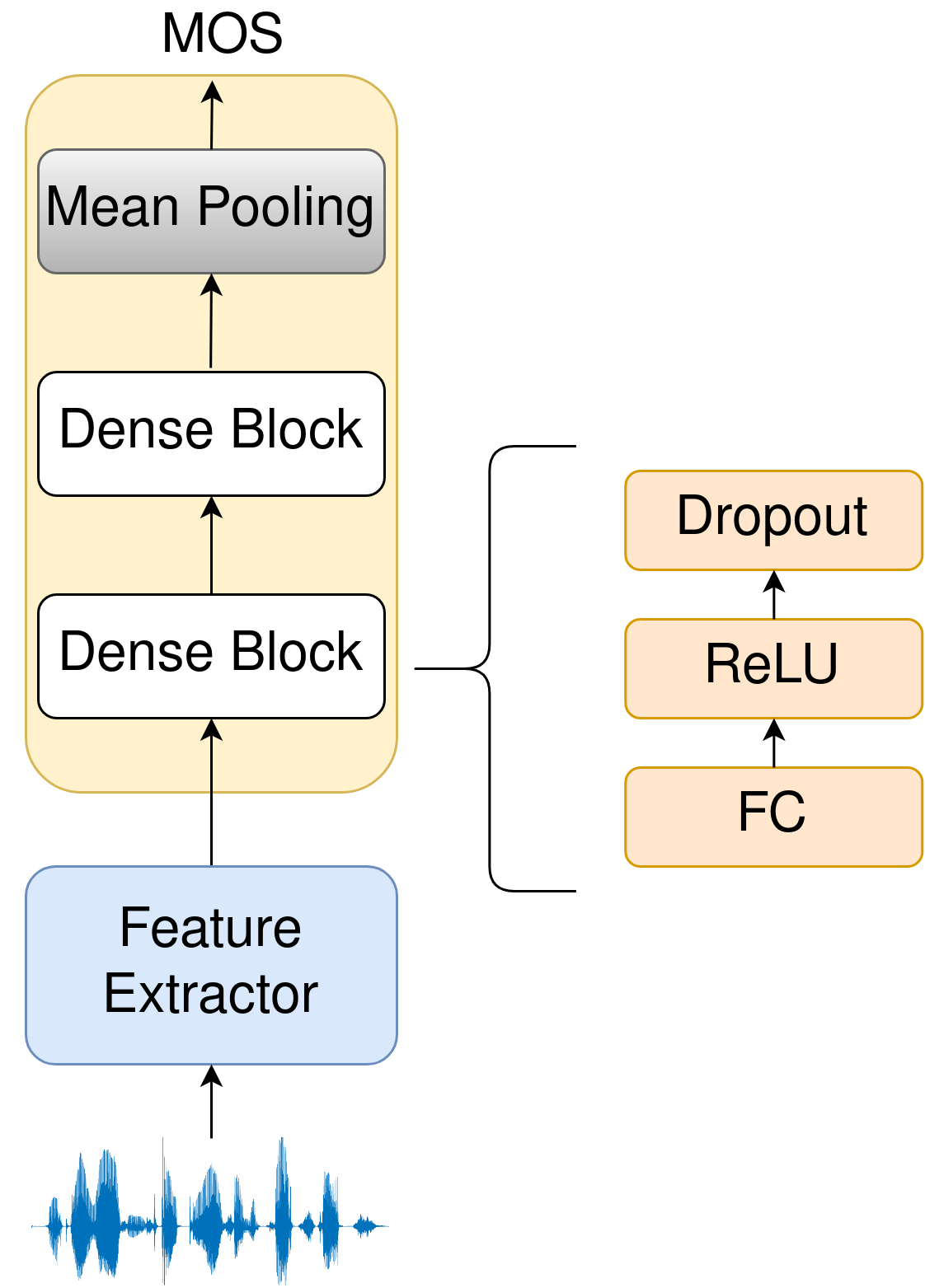}
\caption{The proposed model consists of two modules: Feature Extractor (in blue) and
MOS Predictor (in yellow).}\label{mos_predictor_fc_paper}
\end{figure}

\subsection{Speaker Verification Models}\label{subsection_speaker_verification_models}

The GE2E, Clova, TitaNet, and SpeakerNet models, originally proposed for SV, were selected for speech feature extraction and are discussed in more detail below:

{\bf GE2E} \cite{Wan2018GE2E} is a model that uses the Generalized End-to-End loss function for training and consists of LSTM layers and a fully connected layer with softmax activation. It extracts the vector of embeddings from the log-mel filterbank energies of each speaker's sentences and computes the centroid of each speaker. The similarity matrix is determined from the centroid of each speaker and the parameters learned during training.

{\bf Clova} \cite{Heo2020clova} is a model based on the ResNet architecture \cite{He2016Resnets} proposed in 2020 for speaker recognition. There are two versions: Q/SAP, lighter and with fewer parameters, and H/ASP, which focuses on the quality of the results. Both versions take log-mel-filterbanks as input and use residual blocks and attentive pooling layers. Version Q/ SAP uses self-attentive pooling. The model was trained with a combination of prototypical and softmax angular loss functions. Version H/ ASP achieved higher accuracy and was selected for use in this work.

{\bf SpeakerNet} \cite{Koluguri2020SpeakerNet} is a model with encoder-decoder architecture proposed in 2020 for speaker recognition and verification. It is based on the QuartzNet \cite{Kriman2020QuartzNet} model and has a statistics pooling layer for intermediate feature extraction. The model is trained with the loss functions cross-entropy and additive angular margin. There are two versions, SpeakerNet-L and SpeakerNet-M, with 7M and 5M trainable parameters, respectively. The SpeakerNet-M version showed better results and is used in this work.

{\bf TitaNet} \cite{Koluguri2022TitaNet} is a model with an encoder-decoder architecture proposed in 2022 for speaker verification tasks. It is based on the ContextNet model and has an initial block, a final block, and intermediate blocks that use time-channel separable convolutional layers and residual connections with squeeze and excitation layers. The model uses an attentive statistics pooling layer to extract temporal-independent intermediate features and a decoder consisting of two linear layers.

\subsection{Self-Supervised Learning based Models}\label{subsection_ssl_models}

SSL models are trained with thousands of hours of unlabeled audio. In this work, the following models were selected: Wav2vec 2.0, HuBERT, and WavLM.

{\bf Wav2vec 2.0} \cite{Baevski2020Wav2vec} was developed for the task of automatic speech recognition, which learns latent representations through a process of masking parts of the audio. A new version, XLSR \cite{Conneau2021XLSR}, was trained on a multilingual dataset consisting of 50,000 hours of recordings in 53 languages. The XLS-R \cite{Babu2021Wav2vecXLS-R} version is the latest and has been trained using over 400,000 hours of recordings in 128 languages.

{\bf HuBERT} \cite{Hsu2021Hubert} learns latent representations of speech through training similar to that of Wav2Vec, along with the K-means algorithm used to discretize the input Mel spectrogram. In this work, two versions of the HuBERT model are used, called Large and xLarge, trained with 60,000 hours of English audio data.

{\bf WavLM} \cite{Chen2021WavLMLS} is a more general version of the HuBERT model that can be used for tasks such as speech separation, speaker diarization, speaker verification, and speech recognition. In this work, two versions were selected for evaluation: Large and Base-Plus.

\subsection{Supervised Learning based Model}\label{subsection_supervised_models}

Radford et al. proposed {\it Web-scale Supervised Pretraining for Speech Recognition} (Whisper) \cite{Radford2022Whisper}, an encoder-decoder model based on Transformer, which maps the audio spectrogram to a sequence of text tokens. Whisper was trained through supervised training with approximately 680,000 hours of labeled audio data in English and other 96 languages, including Brazilian Portuguese. Results show that the Whisper model is robust in different scenarios and outperforms SSL-based models when evaluated on different datasets. In this work, five versions were selected for evaluation: Tiny, Base, Small, Medium, and Large.

\section{Experiments}\label{experiments}

This study evaluates a total of 16 models for predicting speech quality. Four of them are based on SV; seven are based on SSL (versions of Wav2vec 2.0 \cite{Baevski2020Wav2vec}, WavLM \cite{Chen2021WavLMLS} and HuBERT \cite{Hsu2021Hubert}); and five based on SL (versions of Whisper). Table \ref{table_experiments_summary} summarizes the models evaluated in this study. This table shows the dimensions of the output embedding and the total parameters of each model, in order to better compare the models.

\begin{table}[!ht]
\centering
\begin{tabular}{|c c c c c|} 
 \hline
 Category & Model & Version & Output dim & Total param\\ [0.5ex] 
 \hline\hline
 Baseline & MOSNet \cite{Lo2019Mosnet} & - & - & 1,1M \\ 
 \hline     
 \multirow{ 4}{*}{SV} & TitaNet \cite{Koluguri2022TitaNet} & Large & [192] & 25,3M \\ 
				            & SpeakerNet \cite{Koluguri2020SpeakerNet} & Medium & [256] & 5M \\ 
				            & GE2E \cite{Wan2018GE2E} & - & [256] & 1,4M \\ 
				            & CLOVA \cite{Heo2020clova} & H/ASP & [512] & 8M \\ 				             				             
 \hline                           
 \multirow{ 7}{*}{SSL} & \multirow{ 3}{*}{Wav2Vec 2.0 \cite{Babu2021Wav2vecXLS-R} } & xls-r-300m & [1024, T] & 300M \\ 
                            & & xls-r-1b & [1280, T	] & 1B \\      
                            & & xls-r-2b & [1920, T] & 2B \\                    

                          & \multirow{ 2}{*}{WavLM \cite{Chen2021WavLMLS}} &  Base-Plus & [768, T] & 94M \\ 
                          & &  Large & [1024, T] & 316M \\  
                          
                          & \multirow{ 2}{*}{HuBERT \cite{Hsu2021Hubert}} & Large & [768, T] & 300M \\ 
                          & & xLarge & [1024, T] & 1B \\  
 \hline
 \multirow{ 5}{*}{SL} & \multirow{ 5}{*}{Whisper \cite{Radford2022Whisper}} &  Tiny & [384, T] & 39M \\ 
                           & &  Base & [512, T] & 74M \\
                           & &  Small & [768, T] & 244M \\    
                           & &  Medium & [1024, T] & 769M \\ 
                           & &  Large & [1280, T] & 1,5B \\ 
 \hline
\end{tabular}
\caption{The MOSNet model is in the {\it Baseline} category; in the SV category, models based on speaker verification; in the SSL category, models based on self-supervised training; in the SL category , models based on supervised training. The "Output dim" column shows the size of {\it embeddings} generated by the {\it Feature Extractor} module. The "Total param" column shows the total set of training parameters for the Feature Extractor module.}\label{table_experiments_summary}
\end{table}

\begin{figure}[!ht]
\centering
\includegraphics[width=4.5cm]{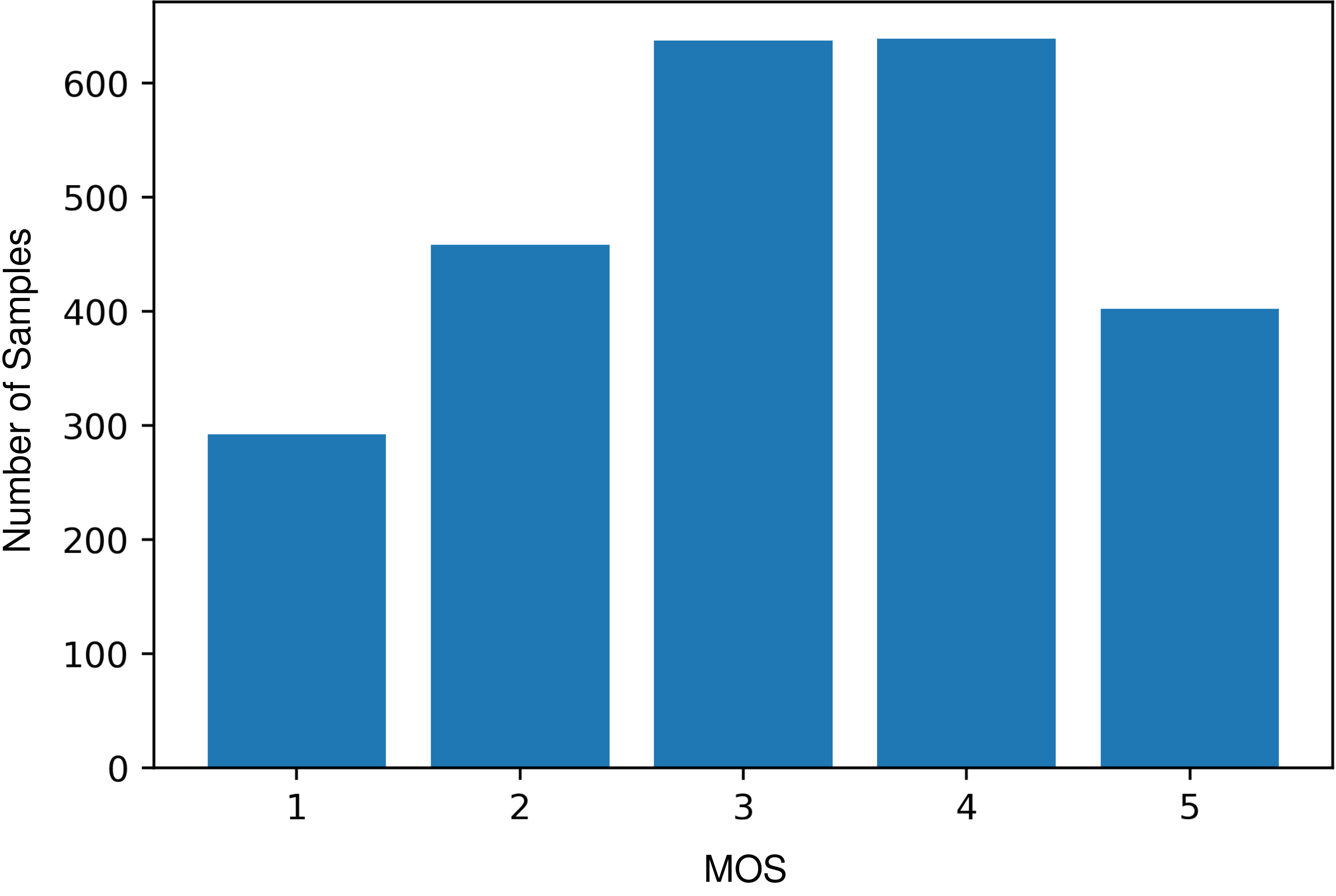}
\caption{Distributions of scores for BRSpeechMOS dataset.}\label{fig_bar_plot_total_samples_score}
\end{figure}

We used two datasets for the experiments in this study: the VCC2018 dataset \cite{Lorenzo2018TheVoiceConversionChallenge} and a Brazilian-Portuguese dataset, which was exclusively created for this present study and is known as BRSpeechMOS. The VCC2018 dataset consists of a total of 28,292 audio samples in English with a sampling rate of 16kHz, each sample being evaluated by 4 evaluators. The BRSpeechMOS dataset contains 2,428 audio samples at 16kHz, and each of these samples has been evaluated by an average of two evaluators. The distribution of scores for the dataset can be seen in Figure \ref{fig_bar_plot_total_samples_score}. This dataset has been utilized to assess the model's performance on a dataset with limited resources.

All evaluated models were first trained with the VCC2018 dataset and then fine-tuned with BRSpeechMOS. Model training was stopped early when no more improvements were observed in a test set, with Spearman correlation analysis. Then, the weights of the best models were selected to be evaluated in a validation set. The experiments were performed on a DGX-1 server running the Linux Ubuntu 18.04 operating system. The server was equipped with a Dual 20-Core Intel Xeon processor E5-2698 v4 2.2 GH, 256 GB RAM, and an NVIDIA® Tesla® V100 GPU.

\section{Results}\label{results}

The results are presented below, grouping the models according to the following categories: speaker verification (SV), self-supervised learning (SSL) and supervised learning (SL). The evaluation metrics used in this study include Pearson correlation (LCC), Spearman rank correlation coefficient (SRCC), Kendall-Tau rank correlation (KTAU), and mean square error (MSE).

\subsection{VCC2018 Experiments}

Table \ref{table_results_val_sv_models} shows the results of all performed experiments using the VCC2018 dataset. For comparison purposes, the results of the experiments using the MOSNet \cite{Lo2019Mosnet} model are also presented. Among the SV models, TitaNet obtained the best results in all metrics, with LCC=0.6933, SRCC=0.6667, KTAU=0.5005 and MSE=0.0160. However, the SpeakerNet, GE2E, and CLOVA models show similar results, all superior to the MOSNet model.

\begin{table}[!ht]
\centering
\begin{tabular}{|c c c c c c c|} 
 \hline
 Category & Model & Version & LCC $\uparrow$ & SRCC $\uparrow$ & KTAU $\uparrow$ & MSE $\downarrow$ \\ [0.5ex] 
 \hline\hline
 \multirow{ 5}{*}{SV} & MOSNet & - & 0.5588 & 0.5159 & 0.3765 & 0.5166  \\ 
                         & TitaNet  & Large (TtN) & {\bf 0.6933} & {\bf 0.6667} & {\bf 0.5005} & {\bf 0.0160} \\ 
                         & SpeakerNet & Medium (SpN) & 0.6428 & 0,6210 & 0.4598 & 0.0202 \\ 
                         & GE2E & - (Ge2) & 0.6118 & 0.5846 & 0,4306 & 0.0193 \\ 
                         & CLOVA & H/ASP (CLO) & 0.6903 & 0.6623 & 0,4966 & 0.0162 \\ 
 \hline
 \multirow{ 7}{*}{SSL} & \multirow{ 3}{*}{Wav2Vec 2.0} & xls-r-300m (Wv3) & 0.7090 & 0.6866 & 0.5190 & 0.0153 \\ 
                          & & xls-r-1b (Wv1) & {\bf 0.7140} & 0.6893 & 0.5210 & 0.0268 \\ 
                          & & xls-r-2b (Wv2) & 0.7014 & 0.6757 & 0.5096 & 0.0159 \\ 
                           
 & \multirow{ 2}{*}{WavLM}  & Base-Plus (WlB) & 0.6917 & 0.6816 & 0.5122 & 0.0163 \\ 
                          & & Large (WlL) & 0.7120 & {\bf 0.7036} & {\bf 0.5316} & {\bf 0.0151} \\ 

 & \multirow{ 2}{*}{HuBERT} & Large (HbL) & 0.6692 & 0.6441 & 0.4800 & 0.0170 \\ 
                          & & xLarge (HbX) & 0.6871 & 0.6684 & 0.5012 & 0.0170 \\  
 \hline
 \multirow{ 5}{*}{SL} & \multirow{ 5}{*}{Whisper} & Tiny (WpT) & 0.7072 & 0.6881 & 0.5187 & 0.0281 \\ 
                          & & Base (WpB) & 0.7178 & 0.6951 & 0.5249 & 0.0225 \\
                          & & Small (WpS) & 0.7136 & 0.6906 & 0.5218 & 0.0212 \\  
                          & & Medium (WpM) & 0.7205 & 0.6957 & 0.5267 & 0.0195 \\ 
                          & & Large (WpM) & {\bf 0.7274} & {\bf 0.7061} & {\bf 0.5365} & {\bf 0.0194} \\ 
 \hline 
 \hline
\end{tabular}
\caption{\label{table_results_val_sv_models}Results of experiments using the VCC2018 dataset.}
\end{table}

Among the SSL models, Table \ref{table_results_val_sv_models} shows that the Wav2Vec 2.0 xls-r-1b model presented the best LCC value, with a value equal to 0.7140. However, in the other metrics, the WavLM-Large model performs best, with SRCC=0.7036, KTAU=0.5316 and MSE=0.0151. On the other hand, HuBERT presented the worst results among the SSL models evaluated. And among the SL models, it appears that the Whisper Large model presented the best results, with LCC=0.727, SRCC=0.7061, KTAU=0.5365 and MSE=\linebreak0.0194. It is worth mentioning that the Whisper Large model presented the best results among all the models using the VCC2018 dataset.

\subsection{BRSpeechMOS Experiments}

Table \ref{table_results_val_sv_models_brspeech} shows the results of all experiments using the BRSpeechMOS dataset. The following experiments using the MOSNet model are also presented: {\it MOSNet ZeroShot} (MZS), which was trained using only the VCC2018 dataset and follows the methodology and hyperparameters used by the original authors; {\it MOSNet From Scratch} (MFS), which was trained exclusively with the BRSpeechMOS dataset; and {\it MOSNet Fine Tuning} (MFT), which was pre-trained with the VCC2018 dataset and fine-tuned with the BRSpeechMOS dataset.

\begin{table}[!ht]
\centering
\begin{tabular}{|c c c c c c c|} 
 \hline
 Category & Model & Version & LCC $\uparrow$ & SRCC $\uparrow$ & KTAU $\uparrow$ & MSE $\downarrow$ \\ [0.5ex] 
 \hline\hline
  \multirow{ 3}{*}{Baseline} & MOSNet & Zero Shot (MZS) & 0.2196 & 0.2107 & 0.1520 & 0.0611  \\ 
                             & MOSNet & From Scratch (MFS) & 0.5090 & {\bf 0.3677} & {\bf 0.2693} & 0.0452  \\ 
                             & MOSNet & Fine Tuning (MFT) & {\bf 0.5118} & 0.3603 & 0.2612 & {\bf 0.0445}  \\ 
  \hline                          
 \multirow{ 4}{*}{SV} & TitaNet & (TtN) & 0.1012 & 0.1177 & 0.0849 & 0.0623  \\ 
                            & SpeakerNet & (SpN)  & {\bf 0.6963} & {\bf 0.6772} & {\bf 0.5173} &  {\bf 0.0311}  \\ 
                            & GE2E & (Ge2) & 0.2655 & 0.2584 & 0.1791 & 0.0704  \\ 
                            & CLOVA & (CLO) & 0.6860 & 0.6755 & 0.5123 & 0.0359  \\ 
 \hline
  \multirow{ 7}{*}{SSL} & \multirow{ 3}{*}{Wav2Vec 2.0} & xls-r-300m (Wv3) & 0.6739 & 0.6593 & 0.5073 & 0.0335 \\ 
                                                        & & xls-r-1b (Wv1) & 0.6539 & 0.6451 & 0.4937 & 0.0477 \\ 
                                                        & & xls-r-2b (Wv2) & 0.6667 & 0.6439 & 0.4959 & 0.0341 \\ 

                        & \multirow{ 2}{*}{WavLM}  & Base-Plus (WlB) & 0.6082 & 0.5936 & 0.4463 & 0.0382 \\ 
                                                   & & Large (WlL) & {\bf 0.6858} & {\bf 0.6831} & {\bf 0.5275} & {\bf 0.0322} \\ 

                        & \multirow{ 2}{*}{HuBERT} & Large (HbL) & 0.5959 & 0.5863 & 0.4407 & 0.0482 \\ 
                                                   &  & xLarge (HbX) & 0.6262 & 0.6214 & 0.4669 & 0.0368 \\ 
 \hline
   \multirow{ 5}{*}{SL} & \multirow{ 5}{*}{Whisper} & Tiny (WpT) & 0.6587 & 0.6240 & 0.4753 & 0.0564 \\ 
                        & & Base (WpB) & 0.6460 & 0.6083 & 0.4645 & 0.0486 \\ 
                        & & Small (WpS) & {\bf 0.6980} & {\bf 0.6968} & {\bf 0.5400} & {\bf 0.0440} \\    
                        & & Medium (WpM) & 0.6904 & 0.6696 & 0.5161 & 0.0534 \\ 
                        & & Large (WpL) & 0.6956 & 0.6852 & 0.5277 & 0.0777 \\ 
 \hline
 \hline
\end{tabular}
\caption{\label{table_results_val_sv_models_brspeech}Results of experiments using the BRSpeechMOS dataset.}
\end{table}

The results of the experiments using the BRSpeechMOS dataset showed that not all models generalize well in a low-resource dataset. Among the SV models, the SpeakerNet model performed best in all metrics evaluated, with LCC=0.6963, SRCC=0.6772, KTAU=0.5173 and MSE=0.0311, followed by the CLOVA model. The TitaNet model was the one that presented the worst results. We believe that the poor performance on the BRSpeechMOS dataset is due to the small dimension of the output embedding, equal to 192 as shown in Table \ref{table_experiments_summary}, which likely causes the embeddings to specialize in the features that differentiate the speakers. Therefore, more training data would be needed for the MOS Prediction module to accurately map the features to the MOS score.

Among the SSL models using the BRSpeechMOS, the Whisper Large model stood out, with LCC=0.6858, SRCC=0.6831, KTAU=0.5275 and MSE=0.0322. This table also confirms that the HuBERT model has lower performance compared to the other models. And among the SL models, it can be seen that the Whisper Small model had the best performance, with LCC=0.6980, SRCC=0.6968, KTAU=0.5400 and MSE=0.0440, followed by the Whisper Large model. 

\section{Discussion}\label{discussion}

When evaluating SV models using the VCC2018 dataset, all models presented good results. That is, using a dataset with a large number of samples, all models proved to be adequate to predict speech quality, with TitaNet presenting the best results. However, when conducting the same experiments with the BRSpeechMOS dataset, which has 2,428 samples, the results showed that the SpeakerNet model can extract more adequate features to evaluate samples quality even when using a much smaller dataset compared to VCC2018.

\begin{figure}[!ht]
\centering
\includegraphics[width=7.5cm]{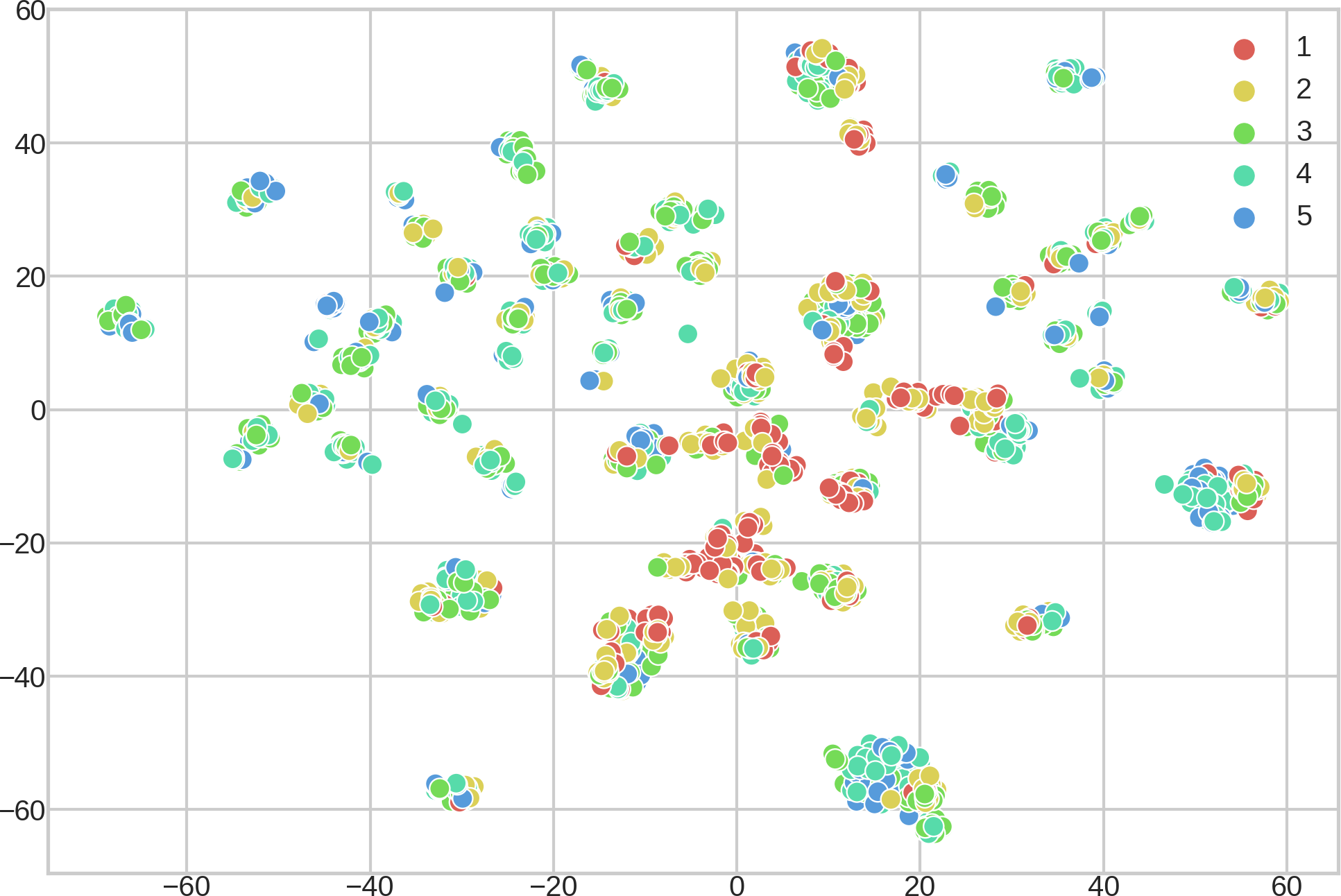}
\caption{T-SNE projection of embeddings from SpeakerNet extracted from BRspeechMOS.}
\label{fig_tsne}
\end{figure}

The representations of the BSpeechMOS using the SpeakerNet model were extracted and projected to 2D space using t-SNE \cite{Van2008visualizing}. Figure \ref{fig_tsne} illustrates the relationship between sample representations and their MOS score. It can be observed that the samples with score 5 (blue), 4 (cyan), and 3 (green) are in clusters. On the other hand, there are also clusters formed by samples with grades 1 (red), 2 (yellow), and 3 (green). Probably, if the BRSpeechMOS samples were evaluated by a larger number of evaluators, the clusters would be more homogeneous. The projections using the VCC2018 are not shown since all the models performed relatively well in the quality prediction task.

Experiments with the SSL models, Wav2Vec 2.0 \cite{Babu2021Wav2vecXLS-R}, WavLM \cite{Chen2021WavLMLS}, HuBERT \cite{Hsu2021Hubert}, and with SL model Whisper \cite{Radford2022Whisper}, using both datasets showed very similar results. The Whisper model showed the best results, which can be justified by a large amount of training data in different languages, including Brazilian Portuguese. In contrast, the HuBERT model showed slightly worse results compared to the other SSL models. This is evidenced by the correlation metrics, as shown in Tables \ref{table_results_val_sv_models} and \ref{table_results_val_sv_models_brspeech}.

\begin{figure}[!ht]
\centering
\includegraphics[width=6cm]{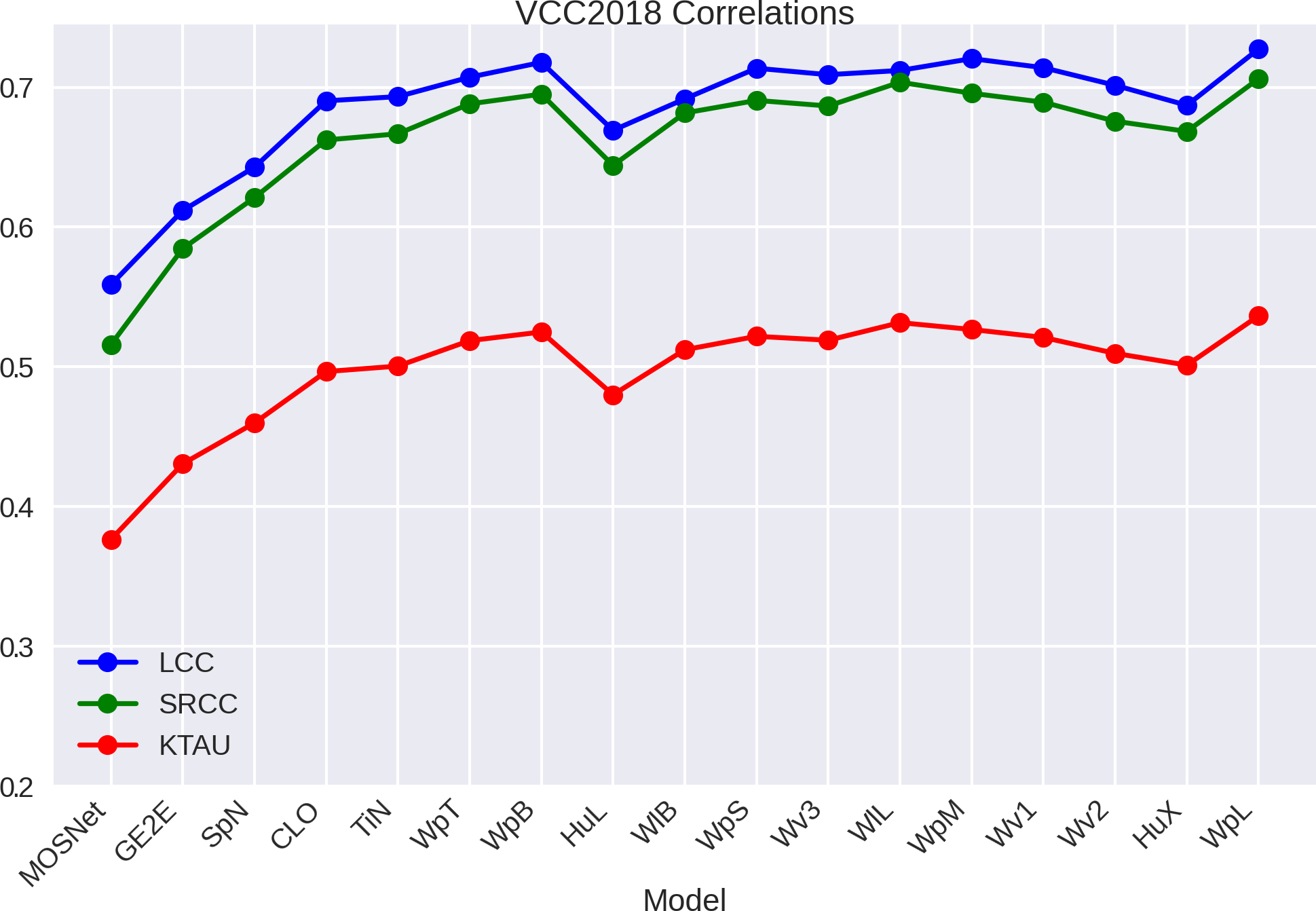}
\includegraphics[width=6cm]{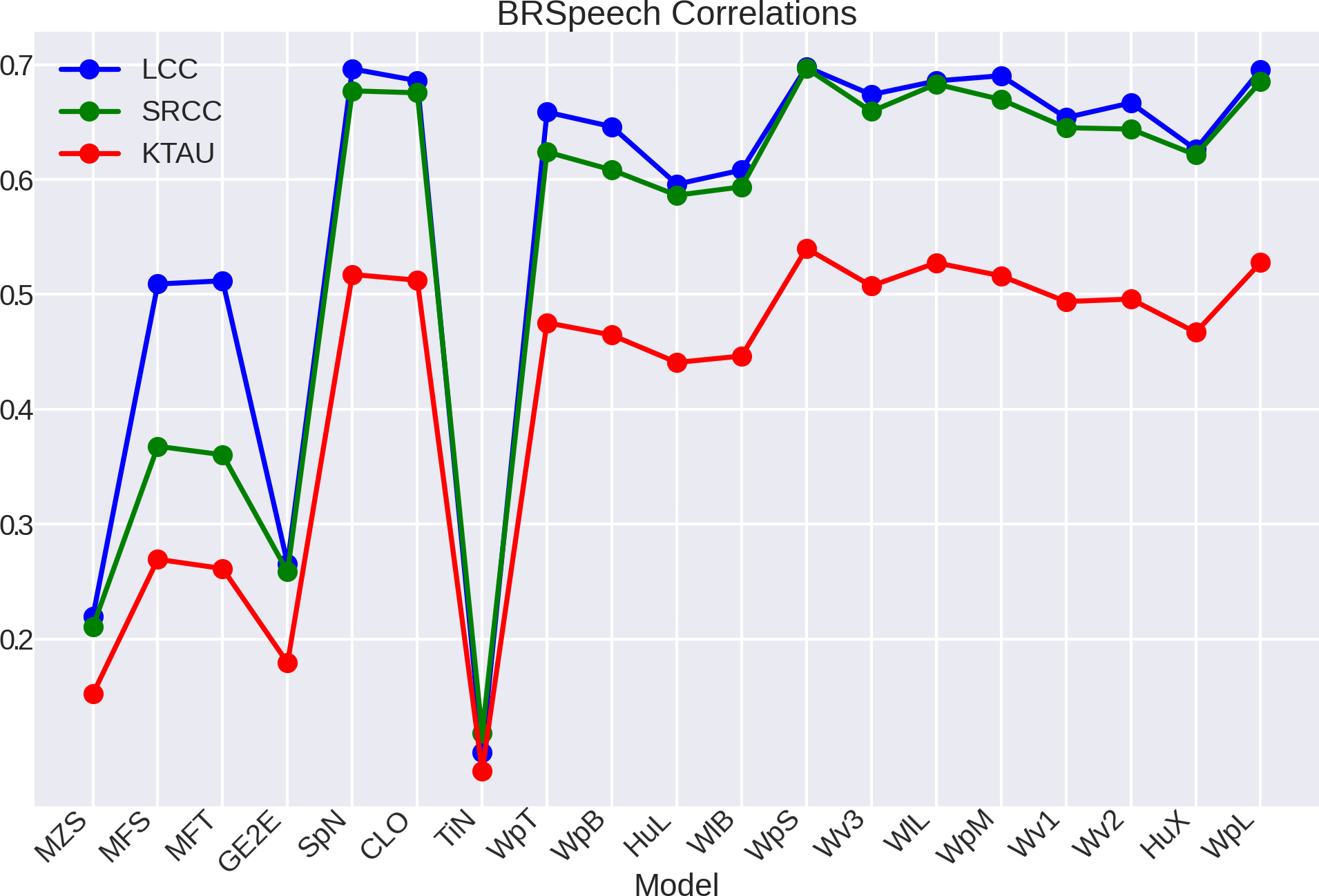}
\caption{On the left, there are graphs displaying the correlation metrics results from the VCC2018 dataset. On the right, there are graphs displaying the results from the BRSpeechMOS dataset. On the x-axis, the models are ordered according to the number of parameters.}
\label{fig_sum_corr_models}
\end{figure}

When comparing all models, it is noticeable that the SL and SSL models are superior to the SV models. However, it is worth noting that the SL model with the best results, SpeakerNet, has only 5M parameters, while the smallest SL-SSL model, Whisper-Tiny, has 39M parameters, almost $8\ times$ the number of parameters of the SpeakerNet model. To better compare the models, Figure \ref{fig_sum_corr_models} shows the ranking of the models with the best results, with the models sorted on the $x$ axis by the number of parameters of the Feature Extractor module.

\section{Conclusions}\label{conclusions}

Our results indicate that the Whisper model, particularly the Large version, is the most effective for the task of speech quality prediction, as demonstrated through its superior performance on the VCC2018 dataset. Additionally, when applied to the BRSpeechMOS dataset, the Whisper model, specifically the Small version, continued to exhibit the highest predictive accuracy, highlighting its ability to generalize well. Furthermore, our study suggests that models designed for speaker verification can also be suitable for predicting speech quality, with the SpeakerNet model performing particularly well, even when using the BRSpeech dataset, which has limited resources and was created exclusively for this study.

\section{Acknowledgements}

The authors are grateful to the Center of Excellence in Artificial Intelligence\footnote{\url{https://ceia.ufg.br/}} (CEIA) at the Federal University of Goias (UFG) for their support and to CyberLabs\footnote{\url{https://cyberlabs.ai}} and Coqui\footnote{\url{https://coqui.ai/}} for their valuable assistance.

% Bibliography
\bibliographystyle{splncs04}
\bibliography{paper}

\begin{thebibliography}{10}
\providecommand{\url}[1]{\texttt{#1}}
\providecommand{\urlprefix}{URL }
\providecommand{\doi}[1]{https://doi.org/#1}

\bibitem{Babu2021Wav2vecXLS-R}
Babu, A., Wang, C., Tjandra, A., Lakhotia, K., Xu, Q., Goyal, N., Singh, K.,
  von Platen, P., Saraf, Y., Pino, J., Baevski, A., Conneau, A., Auli, M.:
  {XLS-R:} self-supervised cross-lingual speech representation learning at
  scale. CoRR  \textbf{abs/2111.09296} (2021),
  \url{https://arxiv.org/abs/2111.09296}

\bibitem{Baevski2022Data2Vec}
Baevski, A., Hsu, W.N., Xu, Q., Babu, A., Gu, J., Auli, M.: Data2vec: A general
  framework for self-supervised learning in speech, vision and language. In:
  International Conference on Machine Learning. pp. 1298--1312. PMLR (2022)

\bibitem{Baevski2020Wav2vec}
Baevski, A., Zhou, H., Mohamed, A., Auli, M.: Wav2vec 2.0: A framework for
  self-supervised learning of speech representations. In: Proceedings of the
  34th International Conference on Neural Information Processing Systems.
  NIPS'20, Curran Associates Inc., Red Hook, NY, USA (2020)

\bibitem{Chen2021WavLMLS}
Chen, S., Wang, C., Chen, Z., Wu, Y., Liu, S., Chen, Z., Li, J., Kanda, N.,
  Yoshioka, T., Xiao, X., Wu, J., Zhou, L., Ren, S., Qian, Y., Qian, Y., Zeng,
  M., Wei, F.: {WavLM}: Large-scale self-supervised pre-training for full stack
  speech processing. IEEE Journal of Selected Topics in Signal Processing
  \textbf{16},  1505--1518 (2021)

\bibitem{Chung2019}
Chung, Y.A., Hsu, W.N., Tang, H., Glass, J.: {An Unsupervised Autoregressive
  Model for Speech Representation Learning}. In: Proc. Interspeech 2019. pp.
  146--150 (2019). \doi{10.21437/Interspeech.2019-1473}

\bibitem{Conneau2021XLSR}
Conneau, A., Baevski, A., Collobert, R., Mohamed, A., Auli, M.: Unsupervised
  cross-lingual representation learning for speech recognition. pp. 2426--2430
  (08 2021). \doi{10.21437/Interspeech.2021-329}

\bibitem{Cooper2022generalization}
Cooper, E., Huang, W.C., Toda, T., Yamagishi, J.: Generalization ability of
  {MOS} prediction networks. In: ICASSP 2022-2022 IEEE International Conference
  on Acoustics, Speech and Signal Processing (ICASSP). pp. 8442--8446. IEEE
  (2022)

\bibitem{Das2020TheVoiceConversionChallenge}
Das, R., Kinnunen, T., Huang, W.C., Ling, Z.H., Yamagishi, J., Yi, Z., Tian,
  X., Toda, T.: Predictions of subjective ratings and spoofing assessments of
  voice conversion challenge 2020 submissions. pp. 99--120 (10 2020).
  \doi{10.21437/VCC\_BC.2020-15}

\bibitem{Fu2018Qualitynet}
Fu, S.W., Tsao, Y., Hwang, H.T., Wang, H.M.: {Quality-Net}: An end-to-end
  non-intrusive speech quality assessment model based on {BLSTM} (2018)

\bibitem{He2016Resnets}
He, K., Zhang, X., Ren, S., Sun, J.: Deep residual learning for image
  recognition. In: 2016 IEEE Conference on Computer Vision and Pattern
  Recognition (CVPR). pp. 770--778 (2016). \doi{10.1109/CVPR.2016.90}

\bibitem{Heo2020clova}
Heo, H.S., Lee, B.J., Huh, J., Chung, J.S.: Clova baseline system for the
  voxceleb speaker recognition challenge 2020. arXiv preprint arXiv:2009.14153
  (2020)

\bibitem{Hsu2021Hubert}
Hsu, W.N., Bolte, B., Tsai, Y.H.H., Lakhotia, K., Salakhutdinov, R., Mohamed,
  A.: {HuBERT}: Self-supervised speech representation learning by masked
  prediction of hidden units. IEEE/ACM Trans. Audio, Speech and Lang. Proc.
  \textbf{29},  3451–3460 (oct 2021). \doi{10.1109/TASLP.2021.3122291},
  \url{https://doi.org/10.1109/TASLP.2021.3122291}

\bibitem{King2016TheBlizzardChallenge}
King, S., Karaiskos, V.: The blizzard challenge 2016 (2016)

\bibitem{Koluguri2020SpeakerNet}
Koluguri, N.R., Li, J., Lavrukhin, V., Ginsburg, B.: Speakernet: 1d depth-wise
  separable convolutional network for text-independent speaker recognition and
  verification (2020). \doi{10.48550/ARXIV.2010.12653},
  \url{https://arxiv.org/abs/2010.12653}

\bibitem{Koluguri2022TitaNet}
Koluguri, N.R., Park, T., Ginsburg, B.: Titanet: Neural model for speaker
  representation with 1d depth-wise separable convolutions and global context.
  In: ICASSP 2022-2022 IEEE International Conference on Acoustics, Speech and
  Signal Processing (ICASSP). pp. 8102--8106. IEEE (2022)

\bibitem{Kriman2020QuartzNet}
Kriman, S., Beliaev, S., Ginsburg, B., Huang, J., Kuchaiev, O., Lavrukhin, V.,
  Leary, R., Li, J., Zhang, Y.: Quartznet: Deep automatic speech recognition
  with 1d time-channel separable convolutions. In: ICASSP 2020 - 2020 IEEE
  International Conference on Acoustics, Speech and Signal Processing (ICASSP).
  pp. 6124--6128 (2020). \doi{10.1109/ICASSP40776.2020.9053889}

\bibitem{Tera2020}
Liu, A.T., Li, S.W., yi~Lee, H.: Tera: Self-supervised learning of transformer
  encoder representation for speech. IEEE/ACM Transactions on Audio, Speech,
  and Language Processing  \textbf{29},  2351--2366 (2020)

\bibitem{Liu2021Tera}
Liu, A.T., Li, S.W., Lee, H.y.: {Tera}: Self-supervised learning of transformer
  encoder representation for speech. IEEE/ACM Transactions on Audio, Speech,
  and Language Processing  \textbf{29},  2351--2366 (2021)

\bibitem{Lo2019Mosnet}
Lo, C.C., Fu, S.W., Huang, W.C., Wang, X., Yamagishi, J., Tsao, Y., Wang, H.M.:
  {MOSNet}: Deep learning-based objective assessment for voice conversion. In:
  Interspeech 2019. {ISCA} (sep 2019). \doi{10.21437/interspeech.2019-2003},
  \url{https://doi.org/10.21437\%2Finterspeech.2019-2003}

\bibitem{Lorenzo2018TheVoiceConversionChallenge}
Lorenzo-Trueba, J., Yamagishi, J., Toda, T., Saito, D., Villavicencio, F.,
  Kinnunen, T., Ling, Z.H.: The voice conversion challenge 2018: Promoting
  development of parallel and nonparallel methods  (04 2018)

\bibitem{Van2008visualizing}
Van~der Maaten, L., Hinton, G.: Visualizing data using t-sne. Journal of
  machine learning research  \textbf{9}(11) (2008)

\bibitem{Oord2018}
Oord, A.v.d., Li, Y., Vinyals, O.: Representation learning with contrastive
  predictive coding (2018). \doi{10.48550/ARXIV.1807.03748},
  \url{https://arxiv.org/abs/1807.03748}

\bibitem{Patton2016Automos}
Patton, B., Agiomyrgiannakis, Y., Terry, M., Wilson, K.W., Saurous, R.A.,
  Sculley, D.: {AutoMOS}: Learning a non-intrusive assessor of
  naturalness-of-speech. CoRR  \textbf{abs/1611.09207} (2016),
  \url{http://arxiv.org/abs/1611.09207}

\bibitem{Radford2022Whisper}
Radford, A., Kim, J.W., Xu, T., Brockman, G., McLeavey, C., Sutskever, I.:
  Robust speech recognition via large-scale weak supervision (2022).
  \doi{10.48550/ARXIV.2212.04356}, \url{https://arxiv.org/abs/2212.04356}

\bibitem{Ragano2022ACO}
Ragano, A., Benetos, E., Chinen, M., Martinez, H.B., Reddy, C.K.A., Skoglund,
  J., Hines, A.: A comparison of deep learning {MOS} predictors for speech
  synthesis quality (2022)

\bibitem{Rix2001Pesq}
Rix, A., Beerends, J., Hollier, M., Hekstra, A.: Perceptual evaluation of
  speech quality ({PESQ}) - a new method for speech quality assessment of
  telephone networks and codecs. In: 2001 IEEE International Conference on
  Acoustics, Speech, and Signal Processing. Proceedings (Cat. No.01CH37221).
  vol.~2, pp. 749--752 vol.2 (2001). \doi{10.1109/ICASSP.2001.941023}

\bibitem{Todisco2019ASVspoof}
Todisco, M., Wang, X., Vestman, V., Sahidullah, M., Delgado, H., Nautsch, A.,
  Yamagishi, J., Evans, N., Kinnunen, T., Lee, K.A.: {ASVspoof} 2019: Future
  horizons in spoofed and fake audio detection. arXiv preprint arXiv:1904.05441
   (2019)

\bibitem{Tseng2021}
Tseng, W.C., Huang, C.Y., Kao, W.T., Lin, Y.Y., yi~Lee, H.: Utilizing
  self-supervised representations for {MOS} prediction. In: Interspeech (2021)

\bibitem{Tseng2022DDOS}
Tseng, W.C., Kao, W.T., Lee, H.y.: {DDOS}: A {MOS} prediction framework
  utilizing domain adaptive pre-training and distribution of opinion scores.
  Interspeech  (2022)

\bibitem{Wan2018GE2E}
Wan, L., Wang, Q., Papir, A., Moreno, I.L.: Generalized end-to-end loss for
  speaker verification. In: 2018 IEEE International Conference on Acoustics,
  Speech and Signal Processing (ICASSP). pp. 4879--4883. IEEE (2018)

\bibitem{Wang2017does}
Wang, S., Qian, Y., Yu, K.: What does the speaker embedding encode? In:
  Interspeech. pp. 1497--1501 (2017)

\bibitem{Wu2019TheBlizzardChallenge}
Wu, Z., Xie, Z., King, S.: The blizzard challenge 2019 (2019)

\bibitem{Yang2022}
Yang, Z., Zhou, W., Chu, C., Li, S., Dabre, R., Rubino, R., Zhao, Y.: {Fusion
  of Self-supervised Learned Models for {MOS} Prediction}. In: Proc.
  Interspeech 2022. pp. 5443--5447 (2022).
  \doi{10.21437/Interspeech.2022-10262}

\bibitem{Zezario2022Mosanet}
Zezario, R.E., Fu, S.W., Chen, F., Fuh, C.S., Wang, H.M., Tsao, Y.: Deep
  learning-based non-intrusive multi-objective speech assessment model with
  cross-domain features. IEEE/ACM Transactions on Audio, Speech, and Language
  Processing  (2022)

\end{thebibliography}

\end{document}